\documentstyle[aps,12pt,preprint]{revtex}

\def \to {\rightarrow}
\def \beq {\begin{equation}}
\def \eeq {\end{equation}}
\def \ba {\begin{eqnarray}}
\def \ea {\end{eqnarray}}
\def \jpsi {J/\psi}

\def \< {\left <}
\def \> {\right >}

\begin{document}
\baselineskip 20pt \renewcommand{\thesection}{\Roman{section}} ~~
{\hfill PKU-TP-97-10(E)} \vskip 20mm

\begin{center}
{\Large {\bf High-$p_T$ $\psi\psi$ production as signals for Double Parton
scattering at hadron colliders}}
\end{center}

\vskip 10mm
\centerline{Feng Yuan}
\vskip 2mm
\centerline{\small {\it
Department of Physics, Peking University, Beijing 100871, People's Republic
of China}}
\vskip 4mm

\centerline{Kuang-Ta Chao}
\vskip 2mm
\centerline{\small {\it
China Center of Advanced Science and Technology (World Laboratory), Beijing 100080,
People's Republic of China}}
\vskip 1mm
\centerline{\small {\it Department
of Physics, Peking University, Beijing 100871, People's Republic of China}} 
\vskip 15mm

\begin{center}
{\bf {\large Abstract}}

\begin{minipage}{140mm}
\vskip 5mm
\par

We present an analysis of the $\psi\psi$ production from  double parton (DP)
sacttering and  single parton (SP) scattering in the large $p_T$ region via
color-octet gluon fragmentation.
We find that at the Tevatron the DP $\psi\psi$ production is at the edge of 
the detectability
 at present, and  at the LHC the DP cross section will
dominate over the SP cross section in the lower $p_T({\rm min})$ region ({\it i.e.},
$p_T({\rm min})<7GeV$).
We also conclude that the color-octet mechanism is of crucial importance to the
double $\jpsi$ production at high energy hadron colliders.

\vskip 5mm
\noindent
PACS number(s): 13.85.Ni, 14.40.Gx
\end{minipage}
\end{center}

\vfill\eject\pagestyle{plain}\setcounter{page}{1} \newpage

The multiple parton (MP) interactions, in which two or more pairs of partons
scatter off each other, might become important parts of the structure of
typical collisions at very high energy hadron colliders, mostly because of
the rapidly increase of the cross section for production of jets with
transverse momentum $p_T\ge p_T({\rm min})\approx 2GeV$. Also, the Double
Parton scattering (DP) processes, where two parton-parton hard scattering
takes place within one $p\bar p$ (or $pp$) collision, have always been a
topic to study possible parton-parton correlations in the proton\cite{dp}.
In experiment, to search for the DP signals, one must subtract the
background signals from the single parton (SP) scattering. Any possible
process to study the DP signals must have a large enough ratio of signal to
background events. This is important to theoretical investigations. The
mostly considered DP process is the production of four high-$p_T$ jets via
double parton scattering within a single hadronic collision \cite{4jet}
($4\to 4$ reactions). In such process, two pairs of jets are produced in the
final states and each pair has equal and opposite transverse momentum.
Various hadron collider experiments have searched for this signature \cite
{afs,ua2,cdf1}. However, the four jets production from the SP processes
($2\to 4$ reactions) has a large background. For example, the CDF
collaboration at the Fermilab Tevatron found evidence that $4\to 4$
reactions contribute only about $5\%$ to the production of four jets with
$p_T\ge 25GeV$\cite{cdf1}. Another sort of the DP signals is the double
Drell-Yan processes\cite{dy}, in which double lepton-pair production may
provide a cleaner signature than that of four-jets process. However, such
processes which are due to quark and anti-quark annihilations are much less
important compared to gluon-induced processes in higher energy hadron
collisions. More recently, the production of one-photon + three-jets as
signals for the DP interactions has been studied\cite{drees-han}, and a
strong signal is observed by the CDF collaboration at the Fermilab Tevatron
\cite{cdf2}.

In this note, we discuss the probability of double $\jpsi$ production at
large $p_T$ as signals for DP scattering.
$\jpsi$ production at hadron colliders is of special significance because it
has extremely clean signature through its leptonic decay modes.
In recent years, $\jpsi$ production at large $p_T$ is well studied both in
theoretical and experimental sectors. $\jpsi$ production at large $p_T$ has
two main sources. One is from $b$ decays, and the other is from so-called
``prompt'' production. The contributions from $b$ decays can be isolated by
using a secondary vertex detector, then the prompt $\jpsi$ production can be
studied in experiment\cite{cdfbdecay}. At high energy hadron colliders,
unlike the Drell-Yan production mostly coming from the quark and antiquark
annihilations, prompt $\jpsi$ production dominantly comes from gluon-gluon
fusion process ($gg\to \psi X$). This gluon-induced process would provide
much larger cross section for the production, and then can be used to study
other interesting physics, such as the DP scattering process. 
By triggering four-$\mu $ final states to measure the double $\jpsi$
production at large $p_T$,
we can search for the DP signals in high energy hadronic collisions.

In \cite{csm}, the authors have calculated $\psi \psi $ production by
considering the single $\jpsi$ production at the leading-order within the
conventional Color-Singlet Model. However, as pointed out by Braaten and Yuan
\cite{frag}, the fragmentation contributions will dominate over those from
leading-order processes at sufficiently large $p_T$, although the
fragmentation processes are of higher order in strong coupling constant
$\alpha _s$. Explicit calculations of the contributions to $\psi $ production
at the Tevatron from fragmentation of gluons and charm quarks revealed that
fragmentation dominates over the leading-order gluon-gluon fusion mechanism
for $p_T$ greater than $6GeV$\cite{explicit}. More recently, the CDF
collaboration have reported their measurement of charmonia production at
large $p_T$. They found a large excess of direct production (excluding the
contribution from $b$ decays and the feeddown from $\chi _c$) both for $\jpsi$
and $\psi ^{\prime }$\cite{cdfbdecay}\cite{cdf3}. The experimental
measurement is a factor of $30\sim 50$ larger than the theoretical
prediction of the Color-Singlet Model. Motivated by this ``surplus''
problem, a new mechanism for heavy quarkonium production at large $p_T$,
named as Color-Octet gluon fragmentation has been proposed\cite{com}, which
is based on the factorization formalism of non-relativistic quantum
chromodynamics (NRQCD)\cite{nrqcd}. In this approach, the production process
is factorized into short and long distance parts, while the latter is
associated with the nonperturbative matrix elements of four-fermion
operators. This factorization formalism provides a new production mechanism,
the color-octet mechanism, in which the heavy quark and antiquark pair is
produced at short distance in a color-octet configuration and subsequently
evolves nonperturbatively into physical quarkonium state. In the past few
years, applications of the NRQCD factorization formalism to $\jpsi$($\psi
^{\prime }$) production at various experimental facilities have been studied
\cite{annrev}.

Single $\jpsi$ production at large $p_T$ may dominantly come from gluon
fragmentation contributions. According to NRQCD factorization formalism, the
gluon fragmentation to $\jpsi$ production can be factorized as,
\ba
\label{expansion}
D_{g\to \jpsi}(z,\mu ^2)=\sum\limits_n d_{g\to n}(z,\mu
^2)\langle {\cal O}_n^{\jpsi}\rangle ,
\ea
where $z$ is the longitudinal momentum fraction carried by the produced
$\jpsi$ in gluon fragmentation, $\mu =2m_c$ is the fragmentation scale.
$d_{g\to n}$ represent the short-distance coefficients associated with the
perturbative subprocesses in which a $c\bar c$ pair is produced in a
configuration denoted by $n$ (angular momentum $^{2S+1}L_J$ and color index
1 or 8). $\langle {\cal O}_n^{\jpsi}\rangle $ are the long distance
nonperturbative matrix elements demonstrating the probability of a $c\bar c$
pair evolving into the physical state $\jpsi$. The short-distance
coefficients $d_{g\to n}$ can be obtained from perturbative calculations in
powers of coupling constant $\alpha _s$. $\langle {\cal O}_n^{\jpsi}\rangle $
consist of two kinds of matrix elements, {\it i.e.}, the color-singlet and
color-octet matrix elements (according to the color index 1 or 8). The
color-singlet matrix elements may be related to the quarkonium radial wave
function or its derivatives at the origin, and may be calculated by
potential models or estimated by leptonic decay widths of quarkonium states.
Whereas the color-octet matrix elements can only be determined by fitting
the theoretical prediction of quarkonium production rates to the
experimental data. The relative sizes of these matrix elements can be
determined by their scale properties with $v^2$ according to the NRQCD
velocity scaling rules, where $v$ is the typical relative velocity of the
heavy quark in the bound state. The fragmentation function in Eq.(\ref
{expansion}) is a double expansion in $\alpha _s$ and $v$.

For $\jpsi$ production in gluon fragmentation, the color-octet matrix
element $\langle {\cal O}_8^{\jpsi}({}^3S_1)\rangle $ is smaller than the
color-singlet matrix element $\langle {\cal O}_1^{\jpsi}({}^3S_1)\rangle $
by a factor of order $v^4$ according to the NRQCD velocity scaling rules.
However, the short-distance coefficient for the color-octet term in Eq.(\ref
{expansion}) is larger than that for the color-singlet term by a factor of
order $1/\alpha _s^2$. Numerical results show that color-octet contributions
are 50 times larger than color-singlet contributions\cite{com}. In the
following calculations, we neglect the color-singlet term in gluon
fragmentation in Eq.(\ref{expansion}), and only consider the color-octet
gluon fragmentation. The leading-order color-octet gluon fragmentation to
$\jpsi$ production gives \cite{com}
\beq
\label{frag} D_{g\to \jpsi}^{(8)}(z,\mu ^2)=\frac{\pi \alpha _s(2m_c)}{24}
\frac{\langle {\cal O}_8^{\jpsi}({}^3S_1)\rangle }{m_c^3}\delta (1-z).
\eeq

In our calculations, the effects of the evolution of gluon fragmentation
function with scale $\mu ^2$ are neglected, which may introduce some error.
However, as argued in \cite{sp}, including evolution would not necessarily
be an improvement, since naive Altarelli-Parisi equations do not respect the
phase-space constraint $D_{g\to \jpsi}(z,\mu ^2)=0$ for $z<m_{\jpsi}^2/\mu
^2 $\cite{ap}.

We plot in Fig.1 the typical Feynman diagrams for double $\jpsi$ production
at large $p_T$ via color-octet gluon fragmentation at the Fermilab Tevatron.
Fig.1(a) is one of the Feynman diagrams of the gluon-gluon fusion processes
for the DP interactions, and Fig.1(b) for the SP interactions. In our
calculations of the DP cross section for the large $p_T$ $\psi \psi $
production, we only consider the relevant single $\jpsi$ production at large 
$p_T$ via color-octet gluon fragmentation. For the double $\jpsi$ production
from the SP interactions, we also calculate the contributions only from
color-octet gluon fragmentation\cite{sp}, which dominate over the
color-singlet contributions both from the leading-order processes\cite{csm}
and other fragmentation processes \cite{explicit} at large $p_T$ due to the
same reason as pointed out in \cite{com}.

In the DP interaction processes, the two partonic interactions occur
independently of each other\cite{dp,dy,drees-han,csm}. So, the cross section
for $\psi \psi $ production from the DP processes can be related to the
cross section for single $\jpsi$ production from the SP processes by 
\beq
\label{dp}
\sigma _{{ DP}}(\psi \psi )\approx \frac{\sigma _{{ SP}
}(\psi )^2}{2\sigma _{eff}},
\eeq
where the effective cross section $\sigma _{eff}$ represents the possible
correlation effects of the parton distributions in the proton (antiproton).
If parton correlations are negligible, $2\sigma_{eff}$ should approximately
be equal to the total inelastic cross section of $44mb$ at the Tevatron\cite{x-inel}.
This implies $\sigma_{eff}\approx 22mb$.
Parton correlations tend to reduce the effective cross section ({\it i.e.},
increase the double parton scattering cross section) relative to the
uncorrelated case.
In the following calculations, we use the value $\sigma _{eff}=14.5{\rm mb}$
according to the measurement by the CDF\cite{cdf2}. In the literature, some
modifications to the above formula have been introduced for the calculation of
the DP cross section, which may more correctly represent the correlations
between the two partons in one proton such as the energy-momentum
conservation effects\cite{drees-han,csm}. In our calculations, we do not
consider these modifications, because they only cause a little change to
the total cross section and can then be neglected\cite{drees-han,csm}.

To calculate the single $\jpsi$ production rate, we consider $q\bar q/gg\to
gg$ and $q(\bar q)g\to q(\bar q)g$ subprocesses, and then gluon
fragmentation to $\jpsi$. In  gluon fragmentation, the input parameters
are taken to be
\beq
m_c=1.5GeV,~~\alpha _s(2m_c)=0.26,~~\langle {\cal O}_8^{
\jpsi}({}^3S_1)\rangle =0.0106 GeV^3.
\eeq
The value of the color-octet matrix element $\langle {\cal O}_8^{\jpsi
}({}^3S_1)\rangle$ follows the fitted value in\cite{benek} by comparing the
theoretical prediction to the experimental data at the Tevatron. We use the
MRS(A) parton distribution functions\cite{mrs} to generate the production
cross section, and set the renomalization scale and the factorization scale
both equal to the transverse momentum of the fragmenting gluon $\mu
=p_T(g)\approx p_T(\psi )$. A pseudorapidity cut of $|\eta (\psi)|<0.6$ was
also performed on the produced $\psi $s. We obtain the integrated cross
section for single $\jpsi$ production (over $p_T$) as a function of the
minimum $p_T(\psi )$. In the calculations of gluon fragmentation, we also
include the contributions from the $\chi _c$ and $\psi ^{\prime }$ feeddowns
through $g\to \chi _c$ and $g\to \psi ^{\prime }$ followed by $\chi _c\to
\psi \gamma $ and $\psi ^{\prime }\to \psi X$. This means that the
calculated cross section is for the prompt production (excluding the
contributions from $b$ decays). The feeddown contributions give the same
$p_T $ distribution of $\jpsi$ and increase the total rate by a factor
$\approx 1.6$. We estimate this factor from the measured fraction of direct
production in the prompt production\cite{cdf3} (which is $64\%$). The
leptonic decay branching ratio $Br(\jpsi\to \mu^+\mu^-)=0.0597$ is also
multiplied in the cross section.

Substituting the integrated cross section for single $\jpsi$ production into
Eq.(\ref{dp}), we can estimate the cross section for double $\jpsi$
production coming from the DP processes. The final results are plotted in
Fig.2 and Fig.3 for the experiments at the Fermilab Tevatron and at the CERN
LHC respectively. The curves represent the integrated cross sections for
double $\jpsi$ production (over $p_T$) $\sigma (p\bar p(p)\to \psi \psi
+X)\times Br(\jpsi\to \mu ^{+}\mu ^{-})^2$ as the functions of the minimum
$p_T(\psi )$. For comparison, we also plot the contributions from the SP
interactions via double gluon fragmentation within a single parton-parton
scattering. We calculate these contributions by using $q\bar q/gg\to gg$
subprocesses followed by the two gluons fragmentation into two $\jpsi$s
\cite{sp}.
Considering the dominance of the gluon-gluon fusion subprocess in single $\jpsi$
production, as a rough estimate, we can also write the SP $\psi\psi$ production cross
section as \cite{sp}
\beq
\label{sp}
\sigma_{SP}(\psi\psi)\approx {1\over 2}\sigma_{SP}(\psi) \times f_{g\to \psi},
\eeq
where $f_{g\to \psi}$ is the gluon fragmentation probability to $\jpsi$
at large $p_T$.
In these two figures, the solid lines represent the contributions from
the DP interactions, and the dotted lines correspond to the contributions
from the SP interactions.

As for the DP processes, suitable kinematic cuts can be used to
detect the signals.
The SP processes, $gg\to \psi\psi$ produce the $\jpsi$ pair back-to-back in
transverse momentum, whereas the DP processes produce the $\jpsi$ pair unrelated
to each other but balanced in $p_T$ by a hard gluon (quark) jet.

From Fig.2, we can see that at the Tevatron the DP cross section is
smaller than the SP cross section in all $p_T({\rm min})$ region.
Prompt single $\jpsi$ production rate at the Tevatron has been measured with 
$p_T(\psi )>4GeV$ and $|\eta (\psi )|<0.6$, and the integrated cross section
is\cite{cdfbdecay}
\beq
\sigma (p\bar p\to \jpsi+X)\times Br(\jpsi\to \mu ^{+}\mu ^{-})\approx 24 
{\rm nb}.
\eeq
Substituting the above value of the single $\jpsi$ production cross section
into Eq.(\ref{dp}), we obtain the double $\jpsi$ production cross section
from the Double Parton processes,
\beq
\frac{\sigma (p\bar p\to \psi \psi +X)\times Br(\psi \to \mu ^{+}\mu ^{-})^2
}{\sigma (p\bar p\to \psi +X)\times Br(\psi \to \mu ^{+}\mu ^{-})}= \frac{20
{\rm fb}}{24{\rm nb}}=0.83 \times 10^{-6}.
\eeq
which shows that there will be about one $\psi \psi $ event among every $10^6$
single $\psi $ events. The fraction of double $\jpsi$ events from the DP
processes to single $\jpsi$ events is an order of magnitude smaller than
that from the SP process\cite{sp} (where it is $7.6\times 10^{-6}$). The
cross section for double $\jpsi$ production from the DP processes is
estimated to be about $20{\rm fb}$ with $p_T(\psi )>4GeV$ and $|\eta (\psi
)|<0.6$. This indicates that the DP $\psi\psi$ signal is at the edge of the
detectability of the Tevatron at present (considering the integrated
luminosity of $100 {\rm pb}^{-1}$ now accumulated by each of the Tevatron
detectors and the possible inclusion of both the $\jpsi \to \mu^+ \mu^- $and
$\jpsi \to e^+e^-$ modes), and future increases in luminosity
could possibly make it measurable.

However, the DP cross section for double $\jpsi$ production is proportional to the square
of the single $\jpsi$ production cross section (see Eq.(\ref{dp})),
whereas the SP cross section for double $\jpsi$ production is
proportional to single $\jpsi$ production cross section (see \cite{sp}).
The relative importance of the DP contributions to $\psi\psi$ production
against the SP contributions will be changed with the single $\jpsi$
production rate.
At high enough energies, the DP contributions may dominate over the SP
contributions because the single $\jpsi$ production rate increases with
the collider energy.
To see this more clearly, we give the explicit expression for the ratio
\beq
   {\sigma_{DP}(\psi \psi) \over \sigma_{SP}(\psi \psi)}\approx
   {\sigma_{SP}(\psi)\over \sigma _{eff}\times f_{g\to \psi}},
\eeq
which is obtained from Eqs.(\ref{dp}) and (\ref{sp}).
At the Tevatron the ratio is about 0.11. But the ratio will
increase as $\sigma_{SP}(\psi)$ increases with the collider energies.
At the LHC, we find the single $\jpsi$ production rate will be over an
order of magnitude higher than that at the Tevatron (see also \cite{lhc}).
So, the DP contributions to the double $\jpsi$ production will be more
important at the LHC, which is shown in Fig.3.
This figure shows that in the lower $p_T({\rm min})$ region ({\it i.e.},
$p_T({\rm min})<7GeV$) the DP contributions dominate over the SP
contributions.
For $p_T({\rm min})=5GeV$, the DP cross section is
\beq
\sigma_{DP}(\psi\psi)=1.45{\rm pb},
\eeq
while the SP cross section is
\beq
\sigma_{SP}(\psi\psi)=0.63{\rm pb}.
\eeq
In the above calculations, we choose the effective cross section at the LHC
as the same as that at the Tevatron.
In fact, the value of $\sigma_{eff}$ may change its value while the collider
energy $\sqrt{s}$ increases.
However, we notice that the total cross section increases slowly as $\sqrt{s}$
increases, ({\it e.g.}, $\sigma_{tot}\approx 100 {\rm mb}$ at the
LHC while $\sigma_{tot}\approx 80 {\rm mb}$ at the Tevatron \cite{pdg}).
So, we expect that $\sigma_{eff}$ will not change much at the LHC.

For a typical integrated luminosity $\sim 10^4 {\rm pb}^{-1}$ at the LHC,
we would expect the order of $10^4$ events of DP $\psi\psi$.
This indicates that we can detect the DP signals and also measure the
$\sigma_{eff}$ to investigate the possible correlations between partons in
the proton.

It should be emphasized that the color-octet production mechanism
is of crucial importance to the double $\jpsi$ production from the DP
interactions at the Fermilab Tevatron and the CERN LHC, just as it is to the
single $\jpsi$ production at Tevatron\cite{com,benek}.
Within the color-singlet model, the DP contribution is much
smaller than that from the SP interaction, {\it i.e.}, $\sigma_{DP}(\psi
\psi)\ll \sigma_{SP}(\psi\psi)$ \cite{csm}.
However, after including the color-octet mechanism, the DP contribution can
dominate over the SP contribution for lower $p_T({\rm min})$ region at
the LHC.
So, double $\jpsi$ production at large $p_T$ can also provide
another important test for the color-octet production mechanism.

In conclusion, we have calculated double $\jpsi$ production at hadron colliders.
We find that the DP $\psi\psi$ production is at the edge of the detectability
of the Tevatron at present, and at the LHC the DP cross section will
dominate over the SP cross section in the lower $p_T({\rm min})$ region
({\it i.e.}, $p_T({\rm min})<7GeV$).
We also find the new production mechanism, {\it i.e.,} the color-octet mechanism
is of crucial importance to the double $\jpsi$ production at high energy hadron
colliders. Therefore, the measurement of the double $\jpsi$ production rate
would provide an important test for both the DP scattering and the
color-octet production mechanism.

\vskip 1cm

\begin{center}
{\bf {\large {Acknowledgments}\ }}
\end{center}

We are grateful to Dr. C.F. Qiao for discussions, and especially to Prof.
H.Y. Zhou for providing us with the computing program and many enthusiastic
discussions. One of us (F.Y.) thanks the staff of the Physics Department
Computer Center (Room 540) for their kind help. This work was supported in
part by the National Natural Science Foundation of China, and the State
Education Commission of China and the State Commission of Science and
Technology of China.


\newpage
\centerline{\bf \large Figure Captions} \vskip 1cm \noindent
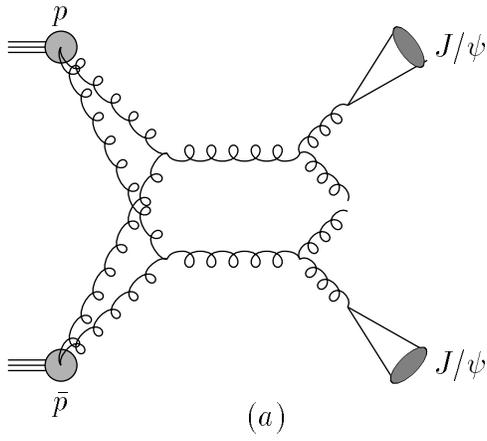
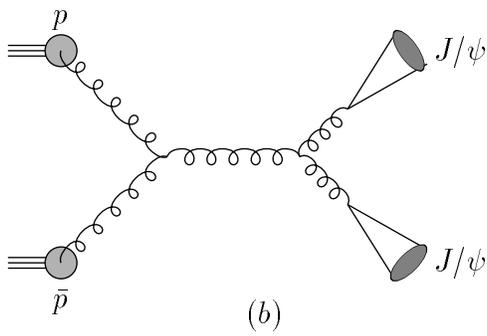
Fig.1. The typical Feynman Diagrams for double $\jpsi$ production at large
$p_T$ via color-octet gluon fragmentation at the Fermilab Tevatron. One of
the diagrams of the gluon-gluon fusion subprocesses for (a) the double
parton interactions, and for (b) the single parton interactions.

\noindent
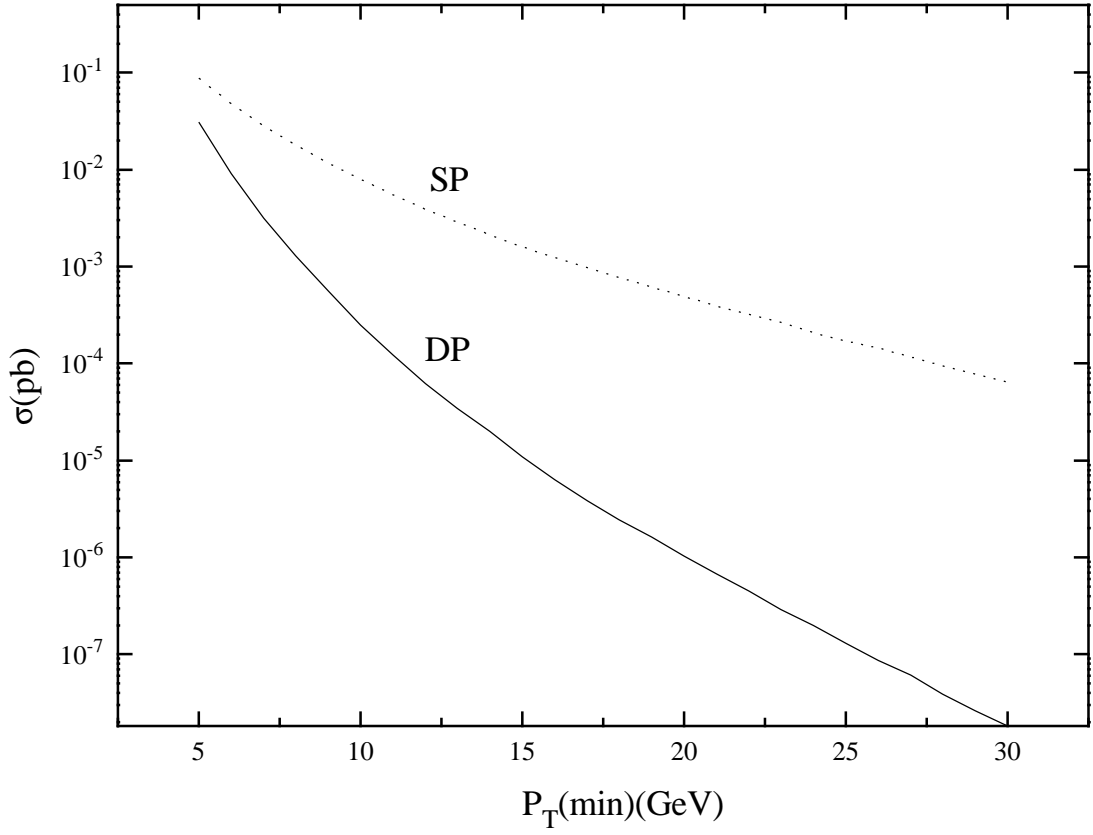
Fig.2. The integrated cross section of $\psi\psi$ production
$\sigma_{\psi\psi}\times Br(\psi \to \mu^+\mu^-)^2$ for $p_T\ge p_T({\rm min}) $
as a function of minimum $p_T(\psi)$ at the Fermilab Tevatron. The
dotted line represents the contributions from the SP processes via double
gluon fragmentation, and the solid line corresponds to contributions from
the DP processes.

\noindent
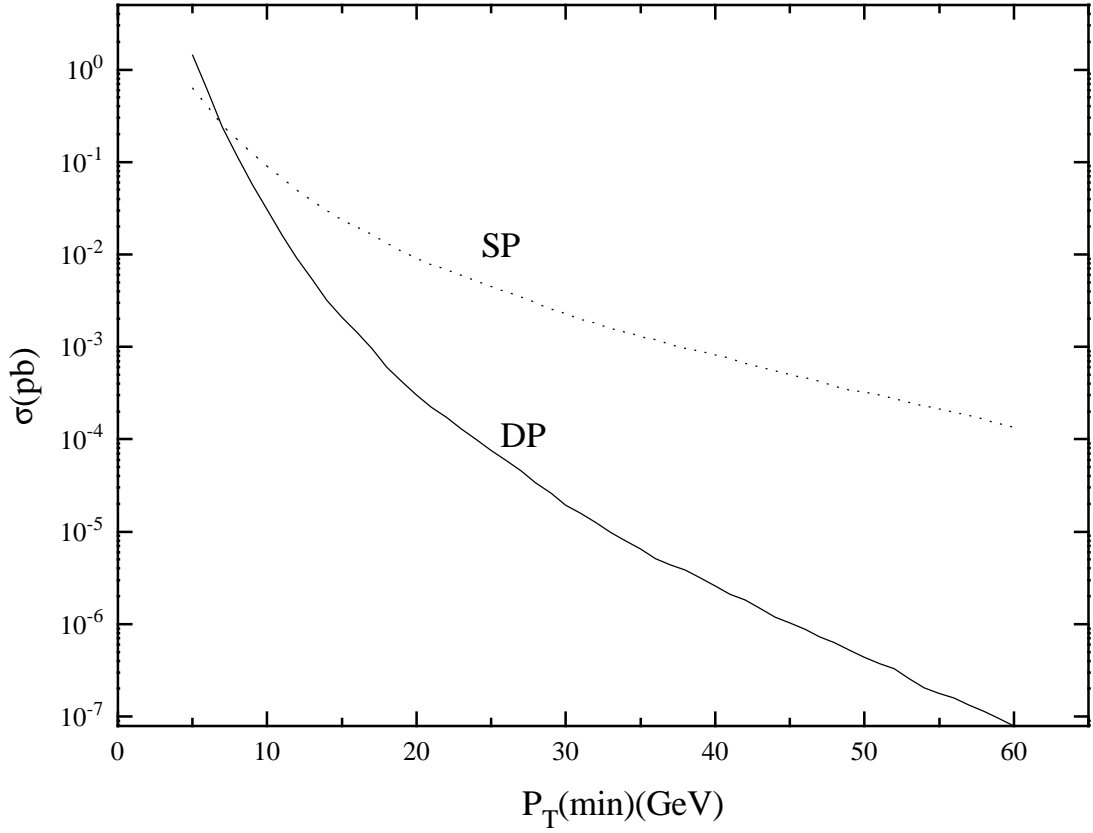
Fig.3. The integrated cross section of $\psi\psi$ production at the CERN
LHC. Here the curves are defined as those in Fig.2.


\begin{references}
\bibitem{dp}  J. Kuti and V. F. Weisskopf, Phys. Rev. {\bf D4}, 3418 (1971);
        C. Goebel {\it et al.}, Phys. Rev. {\bf D22}, 2789 (1980); B. Humpert, Phys.
        Lett. {\bf B131}, 461 (1983).

\bibitem{4jet}  B. Humpert {\it et al.}, Phys. Lett. {\bf B154}, 211 (1985);
        N. Paver and D. Treleani, Z. Phys. {\bf C28}, 187 (1985); T. Sj\"ostrand and
        M. van Zijl, Phys. Lett. {\bf 188}, 43 (1987); {\it ibid.}, Phys, Rev. {\bf 
        D36}, 2019 (1987).

\bibitem{afs}  AFS collaboration, T. Akesson {\it et al.}, Z. Phys. {\bf C34}
        , 163 (1987).

\bibitem{ua2}  UA2 collaboration, J. Alitti, Phys. Lett. {\bf B268}, 145
        (1991).

\bibitem{cdf1}  CDF collaboration, F. Abe {\it et al.}, Phys. Rev. {\bf D47}
        , 4857 (1993).

\bibitem{dy}  M. Mekhfi, Phys. Rev. {\bf D32}, 2371 (1985); F. Halzen {\it 
        et al.}, Phys. Lett. {\bf B188}, 375 (1988).

\bibitem{drees-han}  M. Drees and T. Han, Phys. Rev. Lett. {\bf 77}, 4142
        (1996).

\bibitem{cdf2}  CDF collaboration, F. Abe {\it et al.}, Report No.
        FERMILAB-PUB-97-083-E, Submitted to Phys. Rev. Lett.

\bibitem{cdfbdecay}  CDF collaboration, F. Abe {\it et al}., Phys. Rev.
        Lett. {\bf 69}, 3704 (1992); Phys. Rev. Lett. {\bf 71}, 2537 (1993); K.
        Byrum, FERMILAB-CONF-94/136-E.

\bibitem{csm}  R. W. Robinett, Phys. Lett. {\bf B230}, 153 (1989); L.
        Bergstr\"om {\it et al.}, Phys. Rev. {\bf D42}, 825 (1990).

\bibitem{frag}  E. Braaten and T. C. Yuan, Phys. Rev. Lett. {\bf 69}, 3704
        (1992).

\bibitem{explicit}  M. A. Doncheski, S. Fleming and M. L. Mangano,
        Proceedings of the workshop on Physics at Current Accelerators and the
        Supercollider, FERMILAB-CONF-93/348-T.

\bibitem{cdf3}  CDF collaboration, F. Abe {\it et al}., Report No.
        FERMILAB-PUb-97-024-E; Report No. FERMILAB-PUB-026-E.

\bibitem{com}  E. Braaten and S. Fleming, Phys. Rev. Lett. {\bf 74}, 3327
        (1995); M. Cacciari, M. Greco, M.L. Mangano and A. Petrelli, Phys. Lett. 
        {\bf B356}, 553 (1995); E. Braaten and T.C. Yuan, Phys. Rev. {\bf D52}, 6627
        (1995).

\bibitem{nrqcd}  G. T. Bodwin, E. Braaten and G. P. Lepage, Phys. Rev. D{\bf 
        51}, 1125 (1995).

\bibitem{annrev}  For a recent review to see E. Braaten, S. Fleming, and T.
        C. Yuan, Annu.\ Rev.\ Nucl.\ Part.\ Sci.\ {\bf 46}, 197 (1996).

\bibitem{sp}  V. Barger {\it et al}, Phys. Lett. {\bf B371}, 111 (1996).

\bibitem{ap}  E. Braaten {\it et al.}, Phys. Lett. {\bf B333}, 548 (1994).

\bibitem{x-inel} CDF collaboration, F. Abe {\it et al}., Phys. Rev. {\bf D 44},
        29 (1991).

\bibitem{benek}  M. Beneke and M. Kr\"amer, Phys. Rev.{\bf D 55}, 5269 (1997)

\bibitem{mrs}  A. D. Martin, R. G. Roberts and W. J. Stirling, Phys. Rev. 
        {\bf D50}, 6734 (1994).

\bibitem{lhc} K. Sridhar, Mod. Phys. Lett. {\bf A11} (1996) 1555;
        M. A. Sanchis-Lozano and B. Cano-Coloma, Report No. hep-ph/9701210.

\bibitem{pdg} Partilce Data Group, Phys. Rev. {\bf D 54}, 1 (1996).

\end{references}
\end{document}